\documentclass[twocolumn]{aastex631}
\usepackage{amsmath,amstext}
\usepackage[T1]{fontenc}
\usepackage{apjfonts}
\usepackage[figure,figure*]{hypcap}
\usepackage{multirow}
\usepackage{subfiles}
\usepackage{graphicx}
\usepackage{xspace}
\DeclareGraphicsExtensions{.pdf,.png}

\newcommand{\Aeos}{A{\sc eos}\xspace}

\defcitealias{Emerick2019}{E19}
\defcitealias{AeosMethods}{B24}

\begin{document}

\title{\textsc{Aeos}: Transport of Metals from Minihalos following Population III Stellar Feedback}

\correspondingauthor{Jennifer Mead}
\email{jennifer.mead@columbia.edu}


\author[0009-0006-4744-2350]{Jennifer Mead}
\affiliation{Department of Astronomy, Columbia University, New York, NY 10027, USA}

\author[0000-0002-8810-858X]{Kaley Brauer}
\affiliation{Center for Astrophysics | Harvard \& Smithsonian, Cambridge, MA 02138, USA}

\author[0000-0003-2630-9228]{Greg L. Bryan}
\affiliation{Department of Astronomy, Columbia University, New York, NY 10027, USA}
\affiliation{Center for Computational Astrophysics, Flatiron Institute, 162 5th Ave, New York, NY 10010, USA}

\author[0000-0003-0064-4060]{Mordecai-Mark Mac Low}
\affiliation{Department of Astrophysics, American Museum of Natural History, New York, NY 10024, USA}
\affiliation{Department of Astronomy, Columbia University, New York, NY 10027, USA}

\author[0000-0002-4863-8842]{Alexander P. Ji}
\affiliation{Department of Astronomy \& Astrophysics, University of Chicago, 5640 S Ellis Ave, Chicago, IL 60637, USA}
\affiliation{Kavli Institute for Cosmological Physics, University of Chicago, Chicago, IL 60637, USA}

\author[0000-0003-1173-8847]{John H. Wise}
\affiliation{Center for Relativistic Astrophysics, School of Physics, Georgia Institute of Technology, Atlanta, GA 30332, USA}

\author[0000-0003-2807-328X]{Andrew Emerick}
\affiliation{Carnegie Observatories, Pasadena, CA 91101, USA}

\author[0000-0003-3479-4606]{Eric P. Andersson}
\affiliation{Department of Astrophysics, American Museum of Natural History, New York, NY 10024, USA}

\author[0000-0002-2139-7145]{Anna Frebel}
\affiliation{Department of Physics and Kavli Institute for Astrophysics and Space Research, Massachusetts Institute of Technology, Cambridge, MA 02139, USA}

\author[0000-0002-9986-8816]{Benoit C{\^o}t{\'e}}
\affiliation{Department of Physics and Astronomy, University of Victoria, Victoria, BC, V8P5C2, Canada}


\begin{abstract}
We investigate how stellar feedback from the first stars (Population III) distributes metals through the interstellar and intergalactic medium using the star-by-star cosmological hydrodynamics simulation, \Aeos.  We find that energy injected from the supernovae of the first stars is enough to expel a majority of gas and injected metals beyond the virial radius of halos with mass $M_{\rm dm}\lesssim10^7$ M$_\odot$, regardless of the number of supernovae. This prevents self-enrichment and results in a nonmonotonic increase in metallicity at early times. Most minihalos ($M_{\rm dm} \gtrsim 10^5 \, \rm M_\odot$) do not retain significant fractions of the yields produced within their virial radii until they have grown to halo masses of $M_{\rm dm} \gtrsim 10^7 \, \rm M_\odot$.  The loss of metals to regions well beyond the virial radius delays the onset of enriched star formation and extends the period that Population III star formation can persist.  We also explore the contributions of different nucleosynthetic channels to 10 individual elements. On the timescale of the simulation (lowest redshift $z=14.3$), enrichment is dominated by core-collapse supernovae for all elements, but with a significant contribution from asymptotic giant branch winds to the s-process elements, which are normally thought to only be important at late times. In this work, we establish important mechanisms for early chemical enrichment, which allows us to apply \Aeos in later epochs to trace the evolution of enrichment during the complete transition from Population III to Population II stars.
\end{abstract}

\section{Introduction}
At redshifts of $20 < z < 30$, mere hundreds of millions of years after the Big Bang, the first stars (Population III, or Pop III) began lighting up the Universe, ending the Dark Ages.  Their contribution to the chemodynamical evolution of the Universe as well as their imprint on observed stars today are two of the major frontiers in astrophysics \citep{FrebelNorris2015,KlessenGlover2023}.

In $\Lambda$-cold dark matter cosmology, Pop III stars are expected to form from metal-free primordial gas that collapses in dark matter minihalos with masses of $10^{5}$--$10^{6}\rm{M}_\odot$ \citep[e.g.][]{Haiman1996-1,Tegmark1997,Bromm1999,Abel2002,Yoshida2003,Kulkarni2021}.  The lack of metals in primordial gas makes H$_2$ rovibrational lines the only viable process for cooling at low temperatures ($T<10^4$ K) \citep{Haiman1996-2,Tegmark1997}, resulting in cooling times longer than the freefall time and, thus, large Jeans masses. Pop III stars, then, are thought to have a top-heavy initial mass function (IMF). Many studies have suggested that Pop III stars were very massive, with masses $10^2$--$10^3 \,\rm M_\odot$ \citep{Abel2002,Bromm2002,Bromm2004,Yoshida2008,Hirano2015,Hosokawa2016}.  However, more recent work has demonstrated that fragmentation from turbulence \citep{Latif2013}, weak magnetic fields \citep{Sharda2020}, and gravitational instabilities in disks \citep{Stacy2010,Clark2011,Greif2011,Greif2012}, as well as suppression of accretion by UV radiation \citep{Hosokawa2011,Hosokawa2016,Hirano2014,Latif2022} can reduce the characteristic mass of Pop III stars to tens of solar masses or lower.  A less-top-heavy IMF is also supported by current observational constraints from the numbers and properties of carbon-enhanced metal-poor stars at [Fe/H] $< -5$, which disfavor a flat Pop III IMF of $10-300 \rm M_\odot$ \citep{deBennassuti2017}.

Regardless of the precise shape of the IMF, Pop III stars were likely too massive to survive until present day.  Although several theoretical studies have demonstrated the ability to form Pop III stars with masses below $0.8 \, \rm M_\odot$ \citep[e.g.][]{Jaura2022,ShardaMenon2025,Lake2025}, the non-detection of Pop III stars in Local Group ultrafaint dwarf galaxies (UFDs) thus far seems to reject this possibility. Modeling from \citet{Hartwig2015} and \citet{Magg2019} has shown that given the non-detection of metal-free stars, the probability of Pop III survivors in our stellar halo is unlikely.  \citet{Rossi2021} also demonstrate, by simulating the color-magnitude diagram of Bo{\"o}tes I and comparing to 96 stars with metallicity measurements, that the Pop III $m_{\rm min}>0.8 \, \rm M_\odot$.   Additionally, semi-analytic modeling from \citet{Hartwig2024} has found a best fit $M_{\rm min} = 13.6 \rm M_\odot$.

Even if they do not survive until today, the contribution of Pop III stars to galactic and chemical evolution did not disappear upon their deaths through supernova (SN) explosions; instead, they seeded their surrounding gas, and in some cases, nearby halos \citep[e.g.][]{Smith2015}, with the first metals from which the next generation of stars formed. Among them were slightly metal-enriched low-mass stars (Population II, or Pop II) \citep{Chiaki2016,Chiaki2018,Chiaki2019,Magg2022,Chen2024} that survived to the present \citep{GnedinKravstov2006,Tumlinson2010,Griffen2018}. These stars have retained the chemical fingerprints of Pop III in their atmospheres, which we can observe.  Searching for these rare metal-poor stars is a nontrivial pursuit, but many have been found in the stellar halo as well as across nearly all dwarf satellite galaxies  including UFDs \citep[][]{FrebelNorris2015}. UFDs inefficiently formed stars until being quenched by reionization \citep{Brown2014} and are the ideal cosmic labs in which to study these stars.  Furthermore, UFDs serve as foundational building blocks of larger galaxies, and our own Milky Way's stellar halo is likely comprised of their remnants. The chemical abundances of metal-poor stars in the UFDs and the stellar halo are promising ways to investigate early galaxy formation and chemical enrichment \citep[e.g.,][]{Ji2015,Brauer19}, and stellar chemical abundances have been studied through high-resolution spectroscopy and wide-field spectroscopic surveys such as APOGEE \citep{apogee_overview}, GALAH \citep{GALAH}, LAMOST \citep{LAMOST}, and Gaia RVS \citep{GaiaMission,GaiaDR3}.

A few stars have been found whose abundances suggest they have been enriched by a single progenitor SN \citep[][]{Ezzeddine2019, Skuladottir2021,Yong2021,Xing2023,Skuladottir2024,Ji2024}.  However, our understanding of the progenitors of these stars rests heavily on our assumptions about the explosion mechanisms and yields of Pop III SNe \citep[e.g.][]{HegerWoosley2002,HegerWoosley2010} and the dilution and mixing of metals in the interstellar medium (ISM) and intergalactic medium (IGM).  Thus, having a theoretical understanding of the chemical evolution of the first galaxies, ignited by the first generations of stars, has become critical to moving forward with interpreting the history of our Galaxy.

Cosmological hydrodynamics simulations aimed at studying the chemodynamical evolution of galaxies have been largely successful at reproducing galaxy-scale relations such as the mass-metallicity relation \citep{Obreja2014,Ma2016,Dave2017,DeRossi2017,Hirai2017-2,Torrey2019,Agertz2020,Marszewski2024}, metallicity distribution functions, and the evolution of individual chemical abundances \citep{Marcolini2008,Revaz2009,Sawala2010,RevazJablonka2012,Jeon2017,Hirai2017-1,Hirai2017-2}. Cosmological simulations cannot directly model small-scale physical processes given the large dynamic range, instead relying on subgrid models to include these phenomena \citep[see][and references therein]{Crain2023}.  Among these is the mixing of chemically inhomogeneous particles and regions. Until recently, the assumption of homogeneous mixing was sufficient to reproduce observed global abundance trends across galaxies, but the influx of copious amounts of chemical abundance data for Milky Way and Local Group stars in the last decade has unveiled new variations in the chemical abundances of coeval and conatal populations of stars \citep{Hill2019,Ji2020,Mead2024} that can only be explained by more complex evolutionary processes.

Recent computational advances have opened the door to more directly modeling smaller-scale physics in cosmological simulations by increasing both mass and spatial resolution, thereby reducing the dependence on sub-grid models.  This has been a critical and necessary advance toward understanding the more complex galaxy formation processes that drive chemical evolution, particularly in the early Universe and dwarf galaxies. These higher-resolution simulations
have begun to model single stars as individual particles \citep{Emerick2019,Gutcke+2021,Lahen2020,Lahen+2023,Andersson2023,Andersson2025}.  This enables the study of individual stellar feedback and how the details of when and where individual stars inject energy and enriched material impact the chemodynamical evolution of their environment.

In this paper, we use \Aeos, a new star-by-star cosmological hydrodynamics simulation \citep[][hereafter \citetalias{AeosMethods}]{AeosMethods}, to study the chemical enrichment of minihalos following Pop III and the first Pop II SNe, and the transport of metals between halos and the IGM.  Theory and simulations have demonstrated that dwarf galaxies can efficiently enrich their circumgalactic medium and the IGM \citep{DekelSilk1986,MacLowFerrara1999,Fragile2004,Muratov2017}, showing that outflows of enriched material are primarily dependent on the mass of the halo and the location of the SNe.  Large halos and halos with spatially dispersed SNe are more likely to retain their metals. Other work has examined the effects of stellar feedback on star formation specifically in low-metallicity environments \citep[e.g.][]{Valentini2023}. However, studies of these outflows of enriched material and the effect on subsequent star formation have thus far largely been limited to halos $M_{\rm vir} > 10^6 \, \rm M_\odot$ or low redshift ($z<4$).  Furthermore, studies on the distribution of metals in and beyond host galaxies have been extended to individual elements from different nucleosynthetic origins in theoretical models \citep{KrumholzTing2018} and in hydrodynamic simulations \citep{Emerick2018b}.  These works have found that the mixing and retention of individual elements in a galaxy are dependent on their astrophysical origin.  With \Aeos, we are now able to extend these studies to high redshift minihalos that are the progenitors to dwarf galaxies.

This paper is organized as follows: In Section \ref{sec:sims} we describe the \Aeos model, but we refer the reader to \citetalias{AeosMethods} for additional details.  We explore the behavior of metals in the multiphase ISM and IGM in Section \ref{sec:met_overview} and characterize the loss of metals from halos to the IGM in Section \ref{sec:met_loss}.  In Section \ref{sec:indiv_elem}, we take a look at the individual elements traced in \Aeos, specifically the contributions from different nucleosynthetic sources, and their individual behaviors moving from the ISM to IGM.  Finally, in Section \ref{sec:conclusions}, we summarize our findings, consider the effects of certain choices in our simulation on the robustness of our results, and discuss the implications of our results for future work in numerical and analytical models.

\section{Simulation} \label{sec:sims}
\Aeos is a set of high-resolution cosmological simulations with a box size of 1 comoving Mpc per side using the same initial conditions as \citet{Skinner2020} that treat star particles as individual stars sampled from an assumed IMF, or \textit{star-by-star}, rather than as single stellar populations.  The methods used in these simulations are described in \citetalias{AeosMethods}, and for the case of an idealized galaxy, in \citet[][hereafter, \citetalias{Emerick2019}]{Emerick2019}. The description here is limited to a summary of the properties and physics relevant to this paper.

We follow non-equilibrium primordial chemistry using Grackle \citep{GrackleMethod} with radiative heating and self-shielding from a \citet{HM2012} UV background. To limit the number of radiation sources while still capturing the vast majority of the photon energy budget, we restrict radiation to massive stars with $M_* >$8~M$_{\odot}$. Ionizing radiation from massive stars is tracked using the {\sc Moray} adaptive ray-tracing radiative transfer method \citep{WiseAbel2011}.  Additionally, we trace non-ionizing UV radiation from massive stars that causes photoelectric heating and Lyman--Werner (LW) dissociation of H$_2$.  We track stellar feedback from SNe and stellar winds, and follow yields for 10 metal species (C, N, O, Na, Mg, Ca, Mn, Fe, Sr, Ba) in addition to the total metal density.  

\subsection{Hydrodynamics}
\Aeos uses the adaptive mesh refinement hydrodynamics code {\sc Enzo} \citep{Enzo2014,Enzo2019} to simulate the chemodynamical evolution and feedback of the early universe in high-resolution, with a maximum grid refinement of 1 proper pc.  At this resolution, the Jeans length can become unresolved, leading to artificial fragmentation.  While in \citetalias{Emerick2019}, a pressure floor was used to support gas against artificial fragmentation; \Aeos does not use one, instead setting the star formation density sufficiently low to avoid under-resolving the Jeans length and relying on the efficient conversion of dense gas into stars.

\subsection{Star Formation}

Stars form stochastically if the following conditions are met by assuming that the local star formation rate is proportional to an efficiency per freefall time, which we take as $\epsilon_{\rm ff} = 0.02$.  The conditions for star formation are: (i) $\nabla \cdot v < 0$ (converging gas flow), (ii) $T < 500$ K, and (iii) $n > 10^4$ cm$^{-3}$.

In the simulation, both Pop III and Pop II star formation are treated star-by-star, meaning that each star particle that produces feedback on the timescale of the simulation represents an individual star with a mass sampled from the adopted IMF.

\subsubsection{Population III Star Formation}

In addition to the constraints above, Pop III star formation in our simulation also requires a minimum molecular hydrogen fraction ($f_{H_2} > 0.0005$) that is consistent with high-resolution simulations of Pop III star formation at $n > 10^4$ cm$^{-3}$ \citep{Susa2014}.  To distinguish Pop III from Pop II stars, we use a metallicity limit $Z < 10^{-5} \rm Z_\odot$.

Pop III stellar masses are sampled from a Salpeter IMF \citep{Salpeter1955} with a power-law slope $\alpha = 2.3$ for masses above characteristic mass, $M_{\rm char}$, and an exponential cutoff below \citep[following][]{Wise2012a}. As mentioned in \citetalias{AeosMethods}, there are two runs of the simulation that use different mass ranges and characteristic masses for Pop III.  In this paper, we use the A{\sc eos20} simulation, which has a minimum Pop III mass of 1 $\rm M_\odot$ and maximum mass of 300 $\rm M_\odot$, with $M_{\rm char}$ = 20 $\rm M_\odot$.  Thus, most of our Pop III stars will have masses between $20-300 \rm M_\odot$.  Pop III stars are assigned lifetimes from \citet{Schaerer2002}.

\subsubsection{Population II Star Formation}
Pop II star formation occurs when the metallicity threshold described for Pop III stars is exceeded, and does not require an H$_2$ threshold.  Pop II stellar masses are sampled from a Kroupa IMF \citep{Kroupa2001} with a minimum mass of 0.08 $\rm M_\odot$ and a maximum mass of 120 $\rm M_\odot$.  Stars above 2 $\rm M_\odot$ are treated as individual star particles; however, stars below that threshold in a star forming cell are grouped together into a single particle as they do not contribute to feedback or metal enrichment on the timescales of this simulation.  We set stellar radii, effective temperature, surface gravity, lifetime, and asymptotic giant branch (AGB) phase using the stellar evolution data from PARSEC \citep{Bressan2012}.

\begin{figure}[t!]
    \centering
    \includegraphics[width=\linewidth, trim=1.9cm 2cm 4cm 2cm, clip]{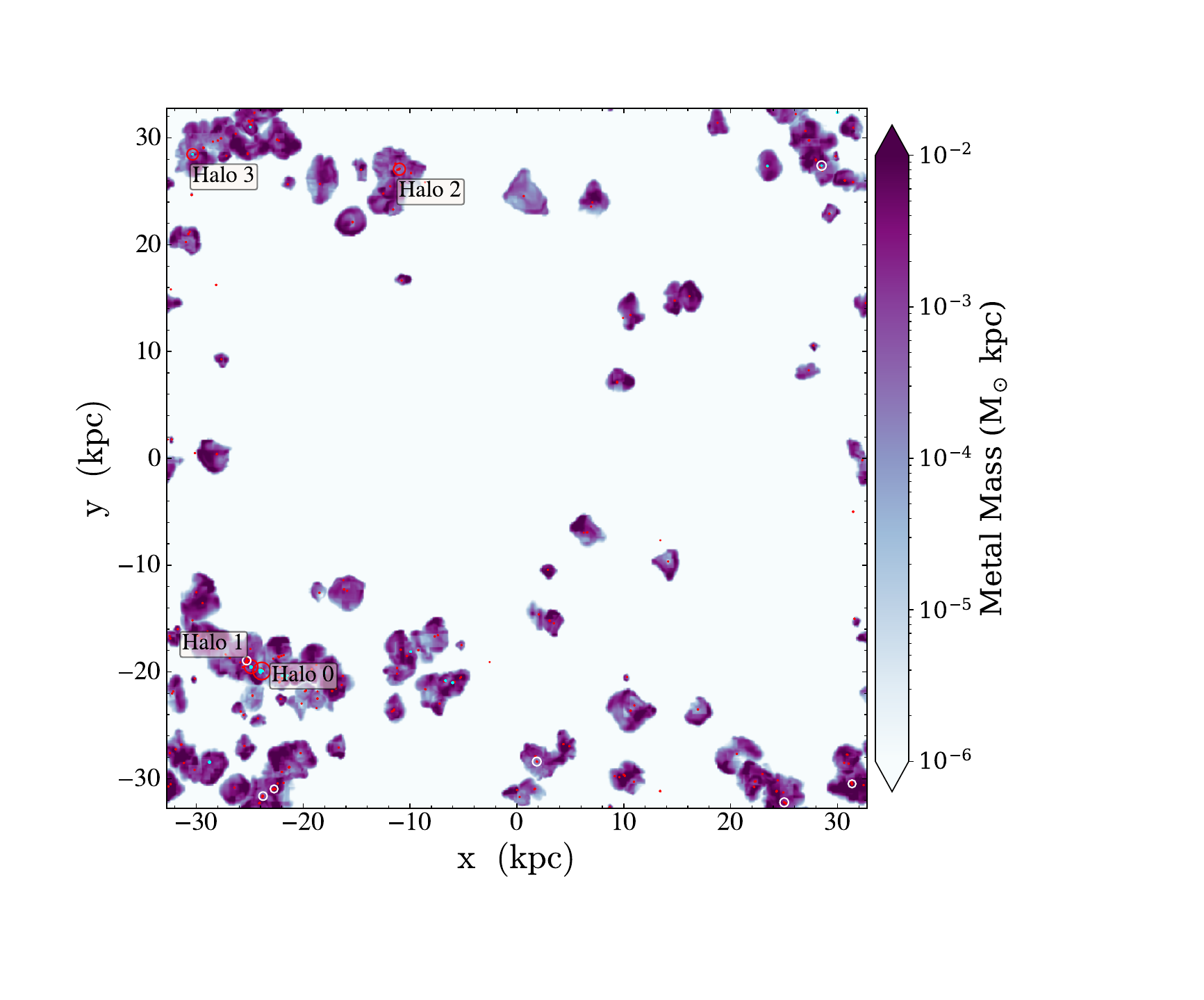}
    \caption{Unweighted projection of metal mass in the \Aeos 1 comoving Mpc box at the final redshift of these runs, $z = 14.26$ ($t = 289.2$ Myr).  The x- and y-axes are in physical units.}
    \label{fig:box_projection}
\end{figure}

\begin{figure*}[t!]
    \centering
    \includegraphics[width=\linewidth]{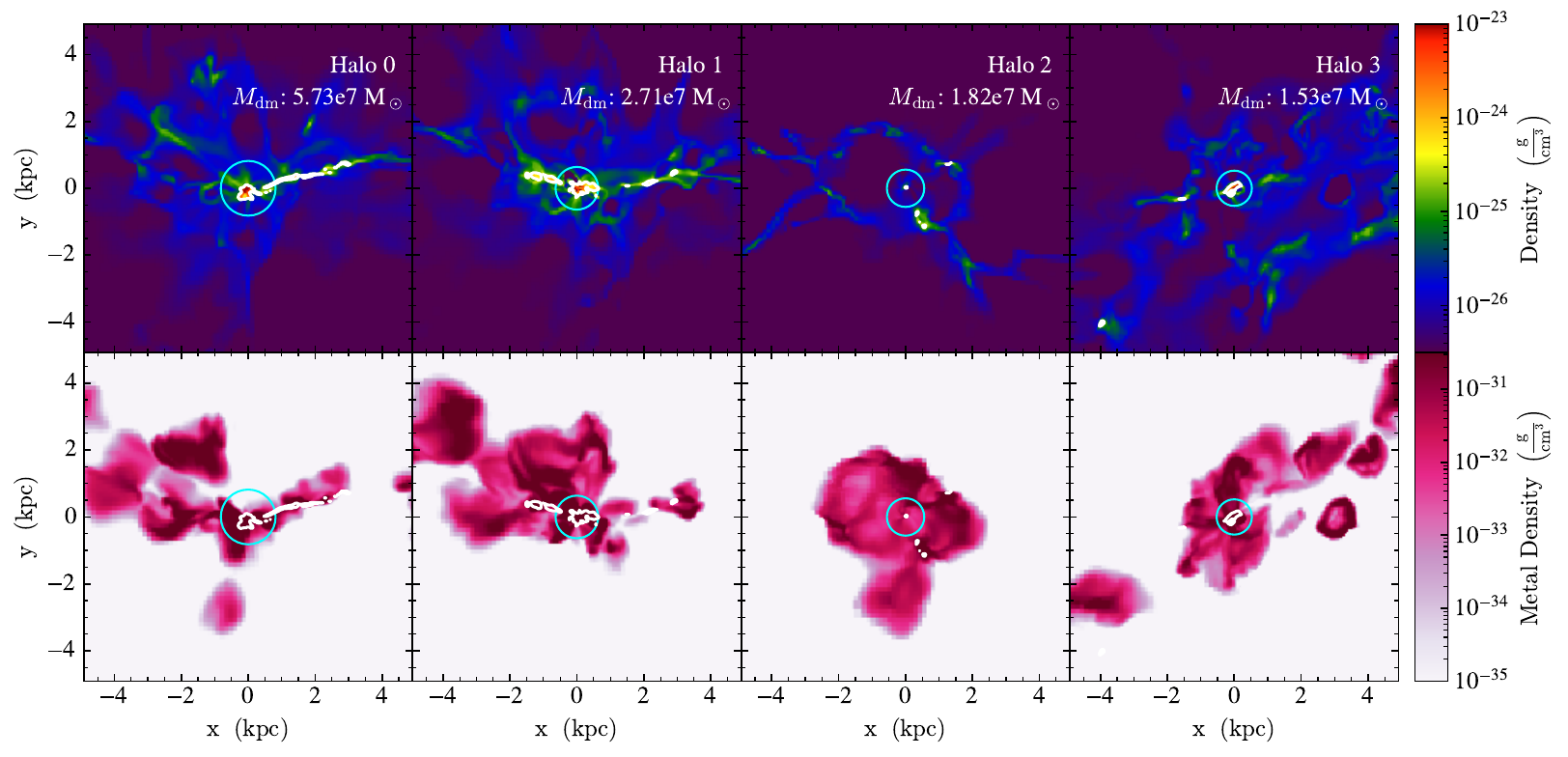}
    \caption{Gas density and metal density slices of the most massive halos in the \Aeos box with $r_{\rm{vir}}$ indicated by the cyan circles.  The white contours represent the $\rho_{200/3}$ density threshold.  The x- and y-axes are in physical units.}
        \label{fig:den_cut}
\end{figure*}

\subsection{Stellar Feedback} \label{sec:feedback}
We implement multi-channel stellar feedback, including core-collapse supernovae (CCSNe) for both Pop III and Pop II stars, Type Ia supernovae (SNIa), and stellar winds for Pop II stars. 
 Each feedback channel is discussed below.

\textbf{\textit{Stellar Winds:}}
There are two types of stellar winds: AGB stellar winds and winds from massive stars. Winds are deposited onto the grid over a 2 pc radius.  As resolving the fast hot winds is computationally expensive (see \citetalias{Emerick2019}), the wind velocity for all massive stars is fixed over their lifetimes at a maximum value of 100 km s$^{-1}$.  The kinetic energy of the winds is fully thermalized before injection.  Mass loss rates are assumed to be constant for all winds over the lifetime of the star.

\textbf{\textit{Supernovae:}}
We implement two different types of SNe: CCSNe and SNIa.  With a maximum resolution of 1 proper pc in these simulations, we are able to resolve the Sedov--Taylor phase of the vast majority of our SNe at the typical gas densities in which they explode \citep{Smith2018b,Hu2019}. Therefore, for these events, we only include the thermal energy deposition.  Energy and mass from each SN explosion are deposited over a 2 pc radius sphere.

\textbf{\textit{Core-Collapse Supernovae:}}
Pop III stars of masses $10 \, \rm M_\odot$ $< M_* \, <$ $100 \, \rm M_\odot$ and Pop II stars of masses $8 \, \rm M_\odot$ $< \, M_* \, <$ $25 \, \rm M_\odot$ explode as CCSNe with a fixed energy of $10^{51}$~erg.  Stars above these ranges undergo direct collapse to a black hole at the end of their lifetimes with no mass or energy feedback.  Although pair-instability SNe are expected to occur for stars in the range of $140-260 \rm M_\odot$, we do not include these in our current simulations due to their rarity and unconfirmed existence.

\textbf{\textit{Type Ia Supernovae:}}
Pop II stars $M_* < 8\, \rm M_\odot$ become white dwarfs, with stars $3 \, {\rm M_\odot} < M_* < 8 \,\rm M_\odot$ being capable of exploding as SNIa. For SNIa, we use the "standard" delay time distribution model in \citet{Ruiter2011}, which sums four different channels: (1) double degenerate, (2) single degenerate, (3) helium-rich donor, and (4) a sub-Chandrasekhar mass.  We use a random number drawn from this distribution to determine when and if a white dwarf will explode as an SNIa.  SNIa explode with a fixed energy of $10^{51}$~erg.

\subsection{Stellar Yields}
We capture detailed chemical yields for Pop III and Pop II CCSNe, SNIa, and stellar winds including C, N, O, Na, Mg, Ca, Mn, Fe, Sr, and Ba metal abundances in addition to H, He, and overall metallicity, Z. The individual metal abundances trace different channels of feedback and progenitor mass.  O, Mg, and Ca primarily trace CCSNe, and trace short timescales of chemical evolution \citep[10 Myr;][]{Timmes1995,Kobayashi2006}. The evolution between SNIa and CCSNe is traced by the relative abundances of O, Mg, and Ca to Fe, which tracks chemical evolution on timescales of 100 Myr to 1 Gyr \citep{Tinsley1980,Matteucci1986,Hayden2015}.  S-process enrichment from low-mass AGB stars is traced using N, Ba (4-8 $\rm M_\odot$), and Sr (\textless 4 $\rm M_\odot$) \citep[e.g.][and references therein]{Karakas2014}.

\textbf{\textit{Pop III CCSNe Yields:}}
Pop III CCSNe ($10 \, \rm M_\odot$ $< \, M_* \, <$ $100 \, \rm M_\odot$) yields are adopted from \citet{HegerWoosley2010} using standard 0.1 mixing.  It is reasonable to approximate that all stars up to 100 $\rm M_\odot$ end their lives as CCSNe, although stars in the range 70-120 $\rm M_\odot$ may end their lives either as CCSNe or by directly collapsing to a black hole \citep{Woosley2017}.

\textbf{\textit{Pop II CCSNe Yields:}}
Yields for Pop II CCSNe were adopted from \citet{Limongi2018} using nine grid points in mass ($13 \, \rm M_\odot$ $\leq \, M_* \, \leq$ $120 \, \rm M_\odot$) and four grid points in metallicity ($3.236\times10^{-5} \leq Z_* \leq 0.01345$).  We use linear interpolation between grid points on the table.  Stars with masses outside of the grid points adopt the abundance ratios of the nearest grid point and scale the yields linearly with mass.  Stars with metallicities outside of the grid points adopt the yields of the closest grid point.

\begin{figure}
    \centering
    \includegraphics[width=\linewidth]{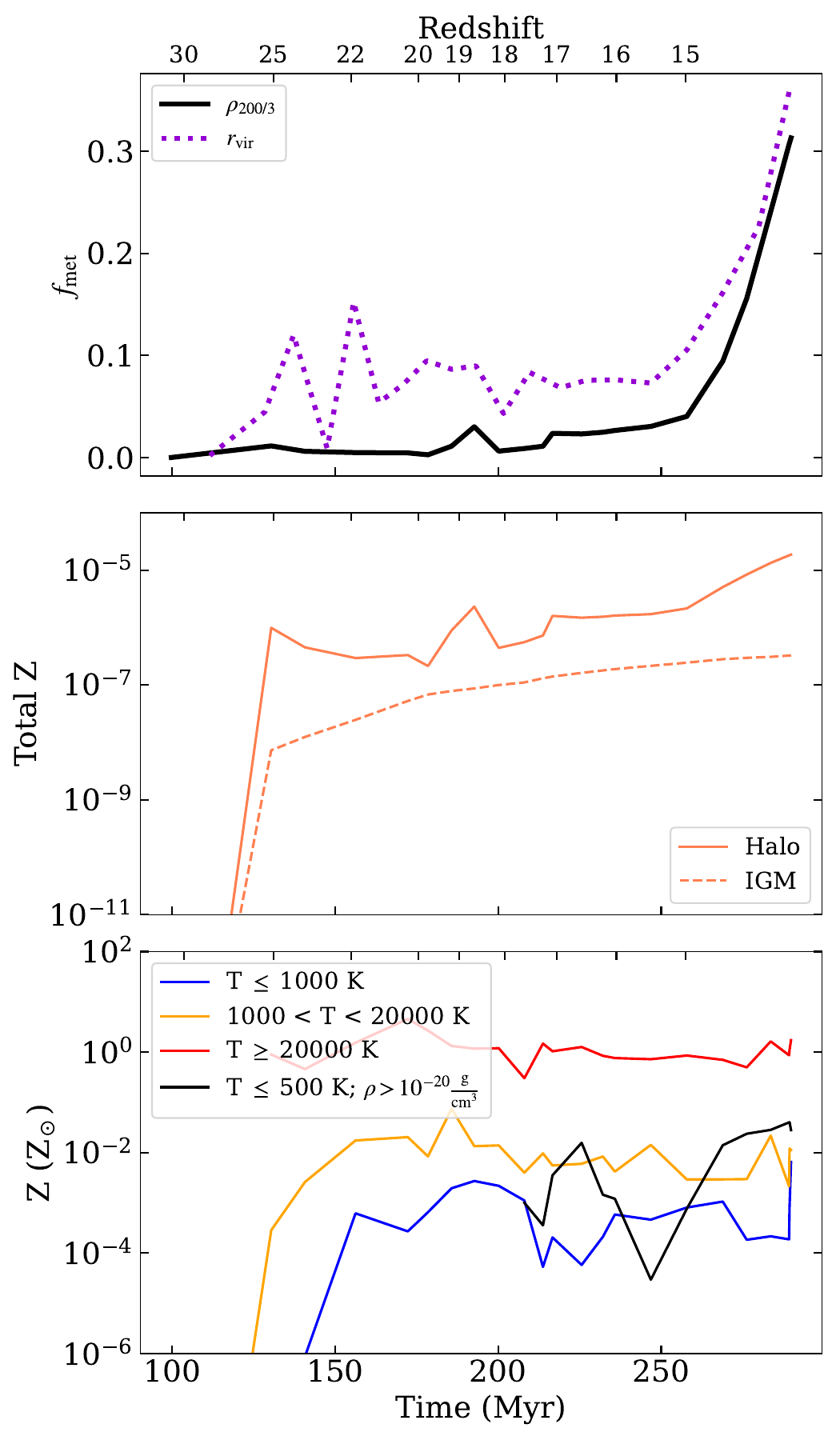}
    \caption{\textit{Top:} Fraction of injected metal mass that remains in halos where the boundary of the halo is either the virial radius {\em (purple)} or the contour where gas is denser than $\rho_{200/3}$, a density that is typical of collapsed halos {\em (black)}.  \textit{Center:} The total metallicities (${M_{\rm met} / M_{\rm gas}}$) of halos {\em (solid)} and the IGM {\em (dashed)} using the $\rho_{200/3}$ density cut. \textit{Bottom:} The average metallicity of hot {\em (red)}, warm {\em (orange)}, and cool {\em (blue)} gas at any density as well as star forming {\em (black)} gas.  Alhough star formation begins much earlier in the simulation, star forming gas does not appear in this plot until 200 Myrs as the data are sparsely sampled in time, and star formation is efficient.  The metallicity of the star forming gas is expected to be no higher than that of cold gas at early times.}
    \label{fig:gas-phase}
\end{figure}

To account for the yields from different stellar rotation models in \citet{Limongi2018}, we adopt a mixture model representing the population-averaged yields using the stellar rotation population fractions in \citet{Prantzos2018}.  Lastly, while these yields reasonably agree with stellar abundance observations in [X/Fe] vs. [Fe/H], Mg is underproduced, so we artificially increase the Mg yield from all massive stars by a factor of 2.2 (see Appendix of \citetalias{AeosMethods}).

\textbf{\textit{Stellar Wind Yields:}}
Yields for AGB winds were adopted from \citet{Cristallo2015} using 8 grid points in mass ($1.3 \, \rm M_\odot$ $\leq \, M_* \, \leq$ $6.0 \, \rm M_\odot$) and 10 grid points in metallicity ($1\times10^{-4} \leq Z_* \leq 0.02$).  The yields for massive stellar winds were also adopted from \citet{Limongi2018} using the same grid points and interpolation as Pop II CCSNe.

\textbf{\textit{SNIa Yields:}}
As discussed in Section \ref{sec:feedback}, the delay time distribution of SNIa is calculated from a combination of four different SNIa sources. We assume a single abundance pattern for all SNIa \citep{Thielemann1986}, which in postprocessing can be rescaled based on the number and type of SNIa that have occurred in the simulation. As a caveat, this assumes that the yields for each SNIa type do not affect the dynamical evolution of our box.

\section{Metals throughout \Aeos} \label{sec:met_overview}
\begin{figure}
    \centering
    \includegraphics[width=\linewidth]{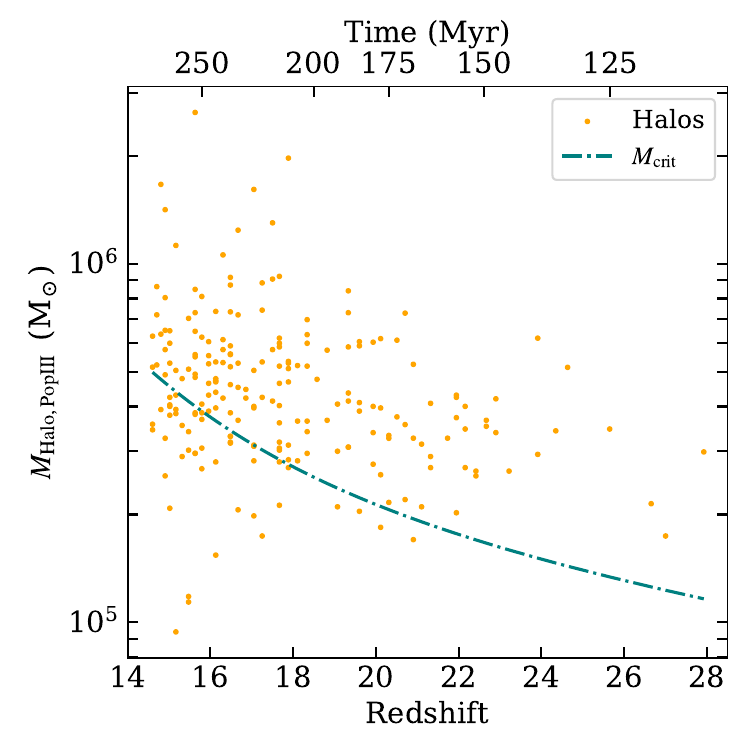}
    \caption{Orange points represent the halo mass at the redshift of first Pop III star formation in a halo's history. The blue dashed line is the theoretical critical mass calculated from \citet{Kulkarni2021}.} 
    \label{fig:first_popiii}
\end{figure}

\begin{figure}
    \centering
    \includegraphics[width=\linewidth]{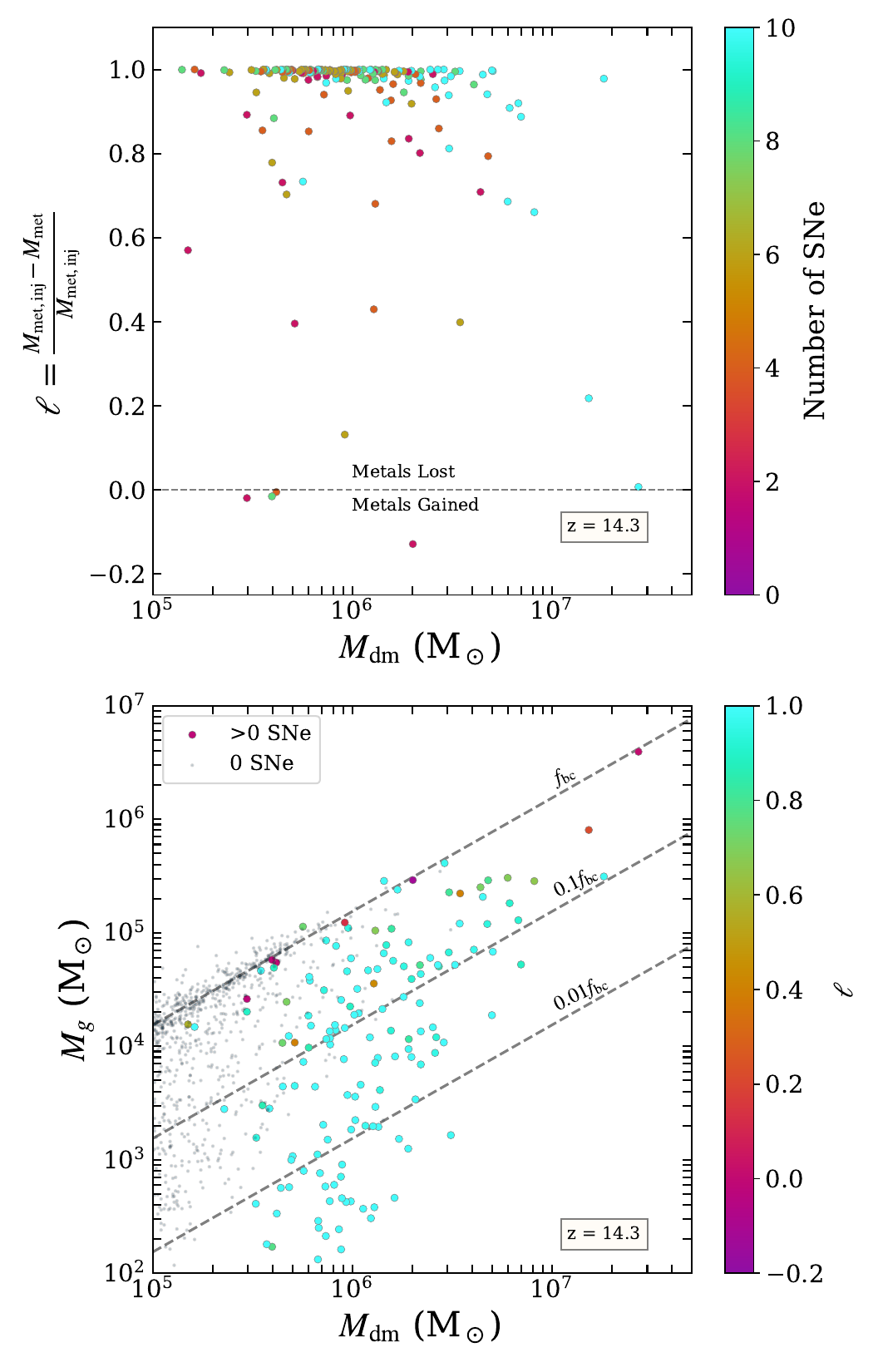}
    \caption{Metal loss fraction, $\ell$. \textit{Top:} Metal loss fraction at the last simulation snapshot at $z=14.3$ colored by the total number of SNe exploded in each halo.  A metal loss fraction of $\ell = 0$ indicates that the current metal mass $M_{\rm met}$ is equivalent to the total metal mass injected into the halo $M_{\rm met,inj}$.  \textit{Bottom:} Gas mass of halos, colored by fractional metal loss at $z=14.3$.  Halos that have not had SNe {\em (black points)} generally have gas masses $M_{g} \sim f_{\rm bc} M_{\rm dm}$, where $f_{\rm bc}$ is the cosmological baryon fraction, $0.154$.  Halos that have had SNe {\em (larger colored points)} have distinctly lower gas masses for the same dark matter masses $M_{\rm dm}$.}
    \label{fig:met_loss}
\end{figure}

The formation of Pop III versus Pop II stars is governed by the metal content of the gas from which they form.  Thus, it is important to gain a broad picture of where metals reside in the simulation.  Metals in \Aeos are tracked both through an overall metal tracer and individual element abundances.  A projection of the unweighted metal mass of the full \Aeos box is shown in Figure \ref{fig:box_projection}.  

To start, we consider the metallicity and fraction of gas phase metals in halos versus what is available in the IGM.  To define the boundary of the halo--IGM interface, we choose to use a density cut at $\rho_{200/3} \, (\equiv \rho_{200}/3)$, as this approximates the density of gas in a virialized region in the absence of feedback \citep{NFW}.  Figure \ref{fig:den_cut} shows examples of gas density and metal density slices of the most massive halos ($M_{\rm dm} > 10^7 \,\rm{M_\odot}$), with both the virial radius (given by Equation \ref{eq:rvir} below) and $\rho_{200/3}$ contour marked. Note that the $\rho_{200/3}$ density contour generally lies well within the virial radius for these halos. From these slices alone, it becomes apparent that a significant fraction of metals lie beyond the boundaries of halos, regardless of how they are defined.

\begin{equation} \label{eq:rvir}
    r_{\rm vir} \approx 526 \, {\rm pc} \left( \frac{M_{\rm vir}}{10^6 \, \rm M_\odot} \, \frac{1}{1+z} \right) ^{1/3}
\end{equation}

The top panel of Figure \ref{fig:gas-phase} shows the fraction of metal mass in halos $f_{\rm met}$, defining the boundary as either the virial radius or the $\rho_{200/3}$ density contour.  Up until roughly 260 Myr ($z \approx 15.3$), $f_{\rm met}= 0$\%--5\% within the $\rho_{200/3}$ contour, or $f_{\rm met}\approx 8$\% on average within the virial radius, emphasizing that most of the metals produced are expelled out of the halo, and more importantly, out of star forming regions.  It is not until after 260 Myr that metals begin to reaccrete to, or are retained within, the most massive halos following SN explosions. Even then, the metal fraction only reaches $f_{\rm met}\sim$31\%--35\%, however, this increase is rapid and may reach even higher fractions in a short amount of time following the end of the simulation.  Of the metals in all halos, $92\%$ reside in the four most massive halos in the simulation.  For comparison, $55\%$ of gas in star-forming halos resides in the four most massive halos, although only $0.8\%$ of gas resides in any halo in the box.

The middle panel of Figure \ref{fig:gas-phase} shows the total metallicity of gas in halos and in the IGM over time.  Despite the small fraction of metals in halos at any given time, the cumulative metallicity of halos is ultimately higher than that of the IGM due to the continuing presence of large swaths of unenriched gas in the IGM.  This, combined with Figures \ref{fig:box_projection} and \ref{fig:den_cut}, suggests that the metals that are not in halos likely reside near halos and do not pollute a large fraction of the IGM.

Regardless of the location, the availability of metals for star formation significantly depends on the phase of the gas that they are in. The metallicity of star forming gas determines whether the stars that form will be Pop III or Pop II.  The bottom panel of Figure \ref{fig:gas-phase} shows the average metallicity of gas in different temperature phases (hot: ${T \ge 2 \times 10^4}$ K, warm: ${10^3 < T < 2 \times 10^4}$ K, cold: ${T \le 10^3}$ K), as well as the metallicity of star forming gas (${T \le 500}$ K, $\rho > 10^{-20} \mbox{ g cm}^{-3}$).  Figure \ref{fig:gas-phase} demonstrates that the hottest gas is the most metal-rich as it is heated by SNe, and the cold gas is the most metal-poor.  Star forming gas has metallicities similar to that of warm and cold gas from 200 Myrs onwards.  Prior to 200 Myrs, the metallicity of star forming gas is inconsistently captured by the simulation snapshots due to the stochasticity and efficiency of star formation.  However, we expect the metallicity of star forming gas to be no higher than that of the cold gas at early times.

\section{Metal Loss from Halos}
\label{sec:met_loss}

The explosions of the first SNe quickly expel gas and metals from halos too small to retain their material. The fraction of metals retained by or, conversely, lost from each halo and the efficiency of mixing of the remaining metals determines when and how quickly star formation from enriched gas takes place in that halo or elsewhere.

Figure \ref{fig:first_popiii} shows that, at $z\lesssim 20$, the first instances of Pop III star formation occur in halos with masses of $M_{\rm dm} \sim 10^5$--$10^6 \, \rm M_\odot$, while, at $z\gtrsim 20$, the mass range narrows somewhat to $\sim 2$--$5 \times 10^5 \, \rm M_\odot$. These results are consistent with those from \citet{CorreaMagnus2024}, which finds that the first minihalos that form Pop III stars are $M_{\rm dm}\sim 10^5 \, \rm M_\odot$, and following nearby radiative feedback, the mass threshold for first Pop III star formation increases to $M_{\rm dm}\sim 10^6 \, \rm M_\odot$. We also compare this finding with the model for critical halo mass $M_{\rm crit}$ for Pop III formation with $v_{bc} = 0$ and $J_{LW} \sim 0 - 1 \, J_{21}$ from Figure 3 of \citet{Kulkarni2021}.  The \citet{Kulkarni2021} model predicts a similar threshold mass to that demonstrated in \Aeos, as well as a scatter of 0.2 dex in halo masses that is also shown in Figure \ref{fig:first_popiii} at lower redshifts.  Given the short lifetimes of Pop III stars, Figure \ref{fig:first_popiii} suggests that the first Pop III SNe exploded in low-mass minihalos with $M_{\rm dm}<10^6 \, \rm M_\odot$.

To track the availability of metals within a given halo, we calculate the fraction of metals produced in the halo that have been lost:
\begin{equation} \label{eq:met_loss}
    \ell = \frac{{M_{\rm met,inj} - M_{\rm met}}}{{M_{\rm met,inj}}}.
\end{equation}
The metal mass ${M_{\rm met}}$ is the total mass of gas phase metals within the virial radius of the halo at the current time step. The injected metal mass ${M_{\rm met,inj}}$ is the total metal mass injected by SNe and AGB stellar winds within the virial radius of the halo or any of its ancestors.  In other words, it is the mass of metals that should be found within the virial radius of the halo if all metals were retained.

The metal loss fraction of each individual parent halo (halos that are not part of a more massive halo) relative to the halo mass is shown in Figure \ref{fig:met_loss}\footnote{An animation of this plot over time is published online alongside this paper. The animated figure shows the metal loss fraction (top) and gas mass (bottom) evolution of individual halos as they grow in dark matter mass.  As time progresses, more halos with SNe appear on the plot.  In the top plot, halos with $M_{\rm dm} \lesssim 10^7 \rm M_\odot$ generally initially appear with values of $\ell < 1$, but quickly lose metals until $\ell \approx 1$.  Halos with $M_{\rm dm} \gtrsim 10^7 \rm M_\odot$ show progressively decreasing $\ell$.  In the bottom plot, halos without SNe cluster around the cosmic baryon fraction with some scatter, continuously increasing in $M_{\rm dm}$ with time.  Halos with SNe generally appear on the plot with gas mass significantly lower than the cosmic baryon fraction.  As time progresses and halos grow in $M_{\rm dm}$, halos that have had SNe slowly increase in $M_g$ until halos $M_{\rm dm} \gtrsim 3\times10^6$ have gas masses within an order of magnitude of the cosmic baryon fraction.}, color-coded by the number of SN that have exploded in the halo by the final redshift, $z=14.3$. Most halos have lost $\sim$100\% of their injected metals, which appears to be independent of the number of SNe that had previously exploded in that halo.  Those halos that have not lost all metals can be divided into two groups:  
\begin{itemize}
    \item Halos $M_{\rm dm} < 10^7 \, \rm M_\odot$ are transient in this stage, having been caught just as metals are being blown out of the halo.  As demonstrated in the animated plot of Figure \ref{fig:met_loss}, halos below this mass quickly lose 100\% of their metals.
    \item Halos $M_{\rm dm}\sim 10^7 \, \rm M_\odot$ ($T_{\rm vir} \sim 3430 \,\rm K$; $v_c \sim 9.7\, \rm km\,s^{-1}$ at $z = 14.3$) and larger, however, can retain metals and slowly build back up a significant fraction of the metals lost through either reaccretion or direct retention of injected metals. 
\end{itemize}

\begin{figure}
    \centering
    \includegraphics[width=\linewidth]{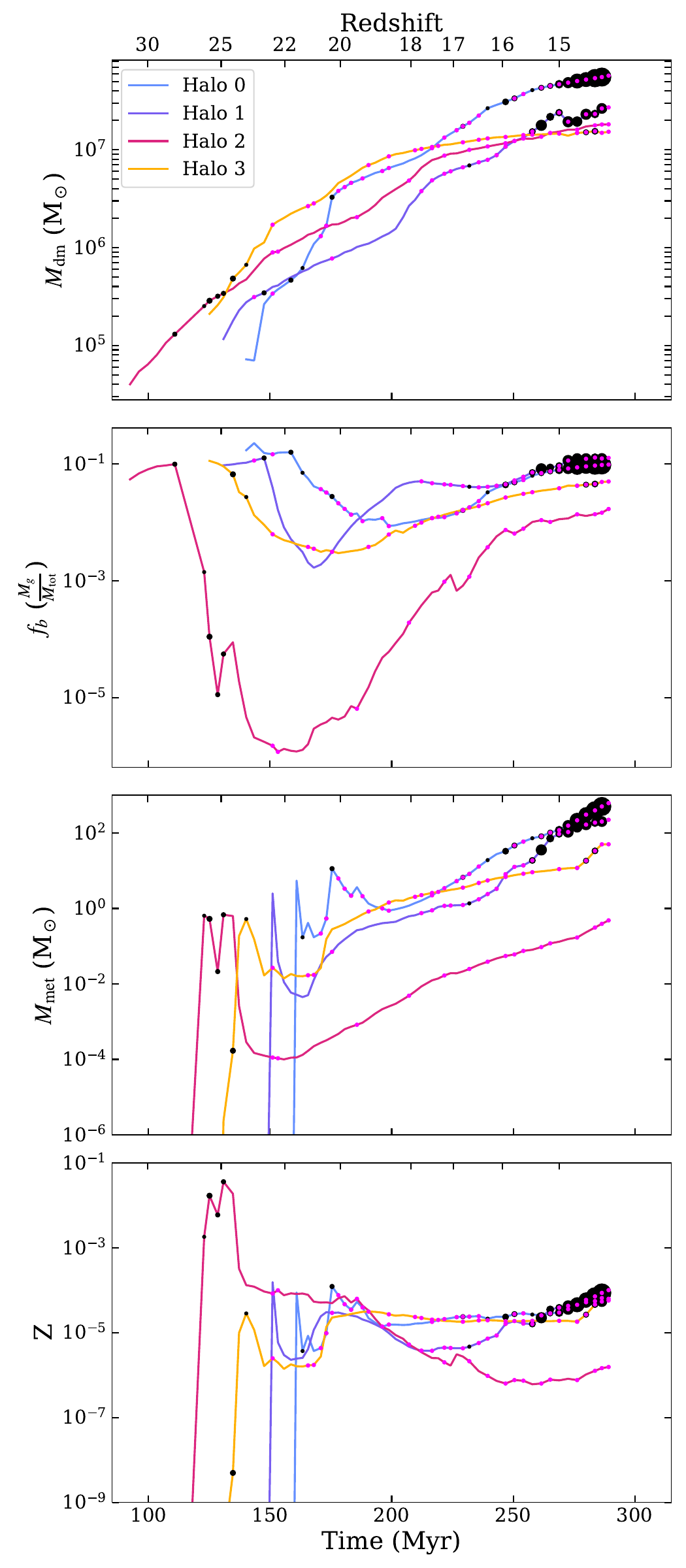}
    \caption{\textit{Top}: Dark matter mass time evolution of the most massive progenitor of the four most massive halos in the simulation.  Black points indicate new SNe in the halo, with size scaled to the number of SNe in that snap shot.  Magenta points indicate merger events. \textit{Top Middle}: Baryon fraction $f_{\rm b} = {M_{\rm g}}/{M_{\rm tot}}$ evolution. \textit{Bottom Middle}: Metal mass time evolution. \textit{Bottom}: Evolution of the total metallicity ($M_{\rm met}/M_{\rm g}$).}
    \label{fig:halo_ev}
\end{figure}

For now, we assert that the reaccretion of metals is a significant contributor to the total amount of metals present in a given halo, which is unsurprising given that most of the metals produced reside in the IGM, as shown in Section \ref{sec:met_overview}.  We also note that there are a handful of halos that have metal loss fractions $\ell < 0$.  These are halos that contain more metals within their virial radius than were locally injected, i.e.\ they are externally enriched (as also discussed in \citetalias{AeosMethods}).  We expect that many halos actually contain metals produced beyond their own virial radius, but this cannot be ascertained by the metal loss fraction alone for halos with $\ell \ge 0$.

In addition to metals, SNe drive out significant fractions of gas from their host halo. The bottom panel of Figure \ref{fig:met_loss} shows the gas mass and dark matter mass of parent halos, color-coded by $\ell$ for halos that have had SNe at the most recent redshift in \Aeos, $z = 14.3$.  The population of halos with SNe are distinct from those without.  The baryon fraction,
\begin{equation}
    f_b = \frac{M_{g}}{M_{\rm dm} + M_{g}} \, ,
\end{equation}
in halos without SNe sits generally around $0.154$, the cosmic baryon fraction, $f_{\rm bc}$.  Halos with SNe diverge strongly from this correlation, with baryon fractions scattered evenly along the range from $0.01f_{\rm bc}$ to $f_{\rm bc}$. The most scatter is found among halos that have lost all of their metals.  

\begin{figure}
    \centering
    \includegraphics[width=\linewidth]{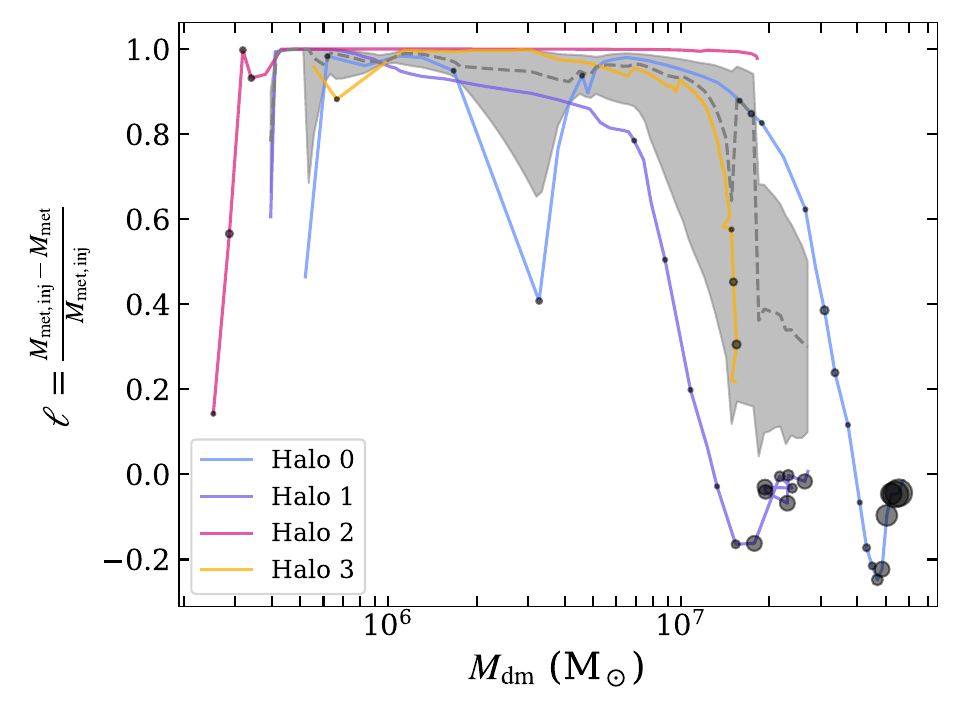}
    \caption{Metal loss fraction of the most massive progenitor line for the four most massive halos in the simulation as a function of $M_{\rm dm}$.  The grey dashed line indicates the median value of $\ell$ for the four most massive halos, while the grey shaded region is bounded by the 16th and 84th percentiles.  Black points indicate new SNe in the halo, with size scaled to the number of SNe in that snapshot.}
    \label{fig:metloss_ev}
\end{figure}

\begin{figure*}
    \centering
    \includegraphics[width=\linewidth]{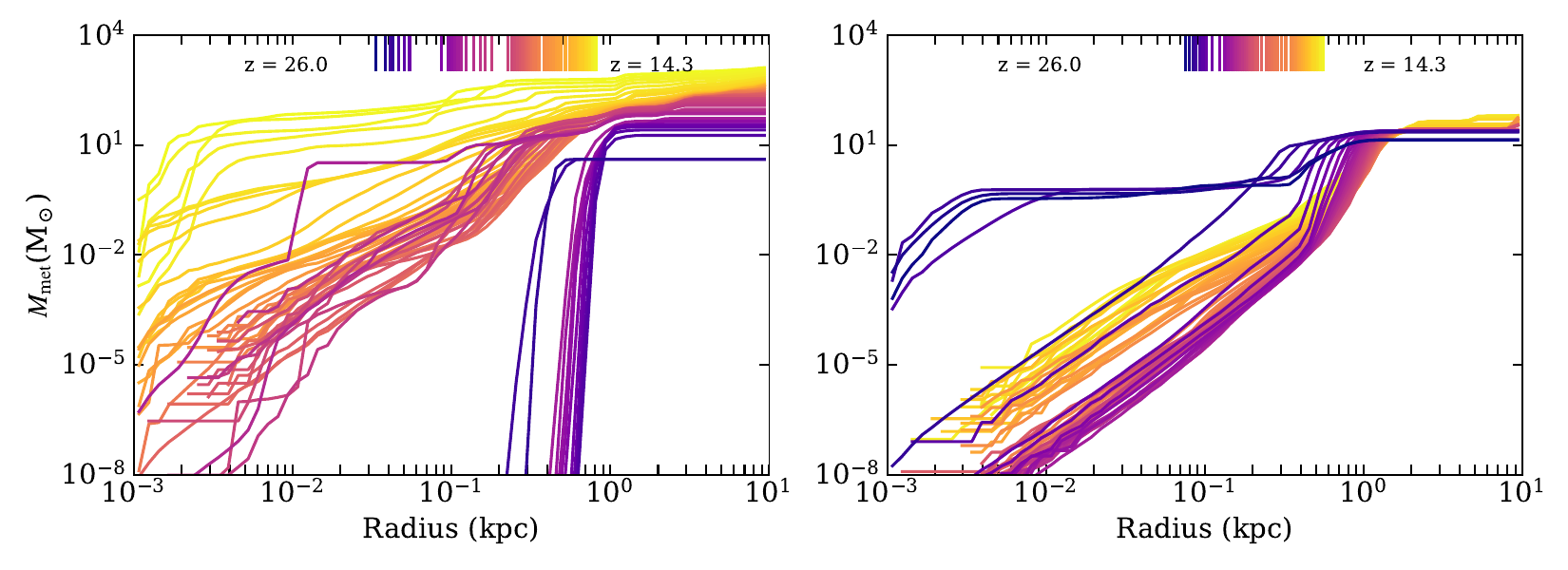}
    \caption{Cumulative metal mass profiles of Halo 0 (left) and Halo 2 (right) in \Aeos over time.  The gradient purple to yellow represents the highest to lowest redshifts respectively.  Lines along the top indicate the virial radius at the redshift for the corresponding profile.}
    \label{fig:met_prof}
\end{figure*}

Halos that have a significant fraction of metals, whether during a transitory stage or once they have reached a mass $M_{\rm dm} > 10^7 \, \rm M_\odot$, also maintain a high baryon fraction, indicating that the loss of metals and gas are linked through the entrainment of metals in gas. Halos with $M_{\rm dm} > 3\times 10^6 \, \rm M_\odot$ ($T_{\rm vir} = 1536 \, \rm K$; $v_c \sim 6.5\, \rm km\,s^{-1}$ at $z=14.3$) begin to reaccrete and retain enough gas to maintain baryon fractions within about an order of magnitude of $f_{\rm bc}$.

\begin{figure*}
    \centering
    \includegraphics[width=0.8\linewidth,trim=0.5cm 1.75cm 2cm 2cm, clip]{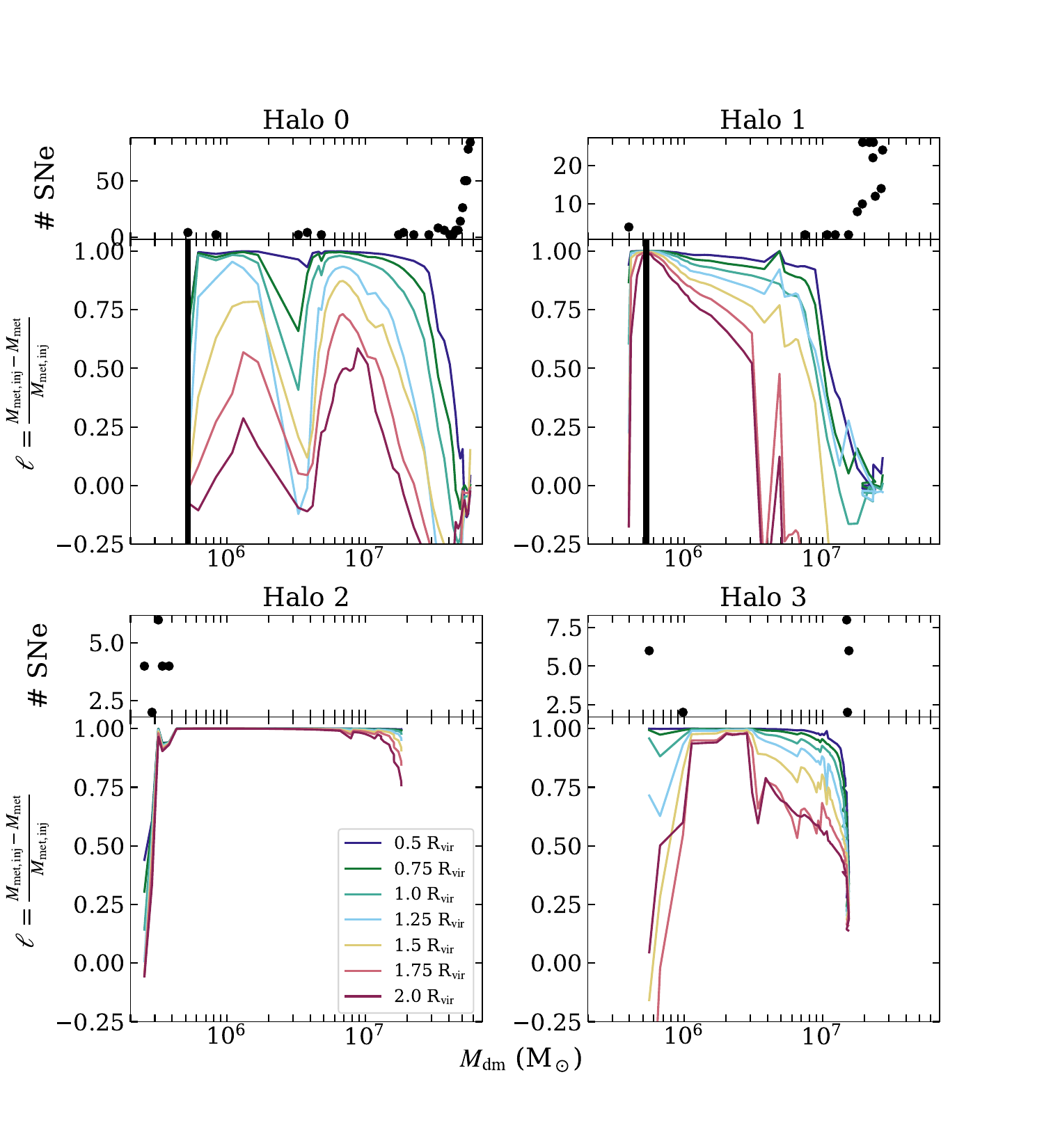}
    \caption{Top panel: Number of SNe in each halo that explode within the virial radius of the halo between snapshots versus $M_{\rm dm}$. 
    Bottom panel: Evolution of the metal loss fraction within different multiples of ${r_{\rm vir}}$. For Halos 0 and 1, the vertical black line depicts the time (160~Myr) at which Halo 0 experienced its first SN explosion, driving metals out of its own halo into the region within 2${r_{\rm vir}}$ of Halo 1.}
    \label{fig:frac_rvir}
\end{figure*}

\begin{figure}
    \centering
    \includegraphics[width=\linewidth]{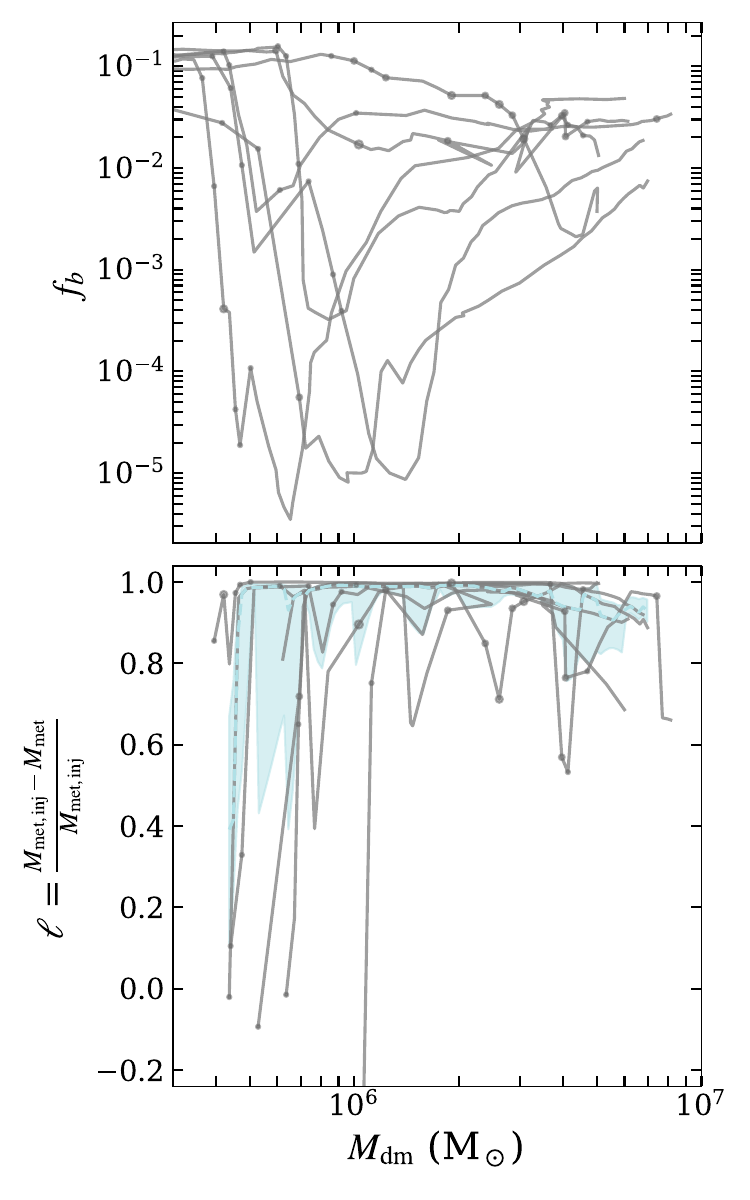}
    \caption{\textit{Top:} Evolution of $f_b$ with dark matter mass of the most massive progenitor of intermediate mass ($5 \times 10^6 < M_{\rm dm} < 10^7 \, \rm M_\odot$) halos. Halos regain their gas $\sim 10^6 \, \rm M_\odot$. \textit{Bottom:} $\ell$ evolution with dark matter mass.  The blue dashed line indicates the median value of $\ell$ for these halos, while the blue shaded region is bounded by the 16th and 84th percentiles. Notably, these halos \textit{do not} regain their metals while at their highest dark matter masses.}
    \label{fig:metloss_ev_int}
\end{figure}

\subsection{Case Study: Massive Halos}
We study the evolution of the most massive halos in our simulation to understand the characteristic evolution of halos that have grown enough to maintain their gas and metals.  There are four halos in our simulation that have dark matter masses $M_{\rm dm} >10^7 \, \rm M_\odot$ at $z = 14.3$.

\subsubsection{Characteristic Evolution}
Figure \ref{fig:halo_ev} shows the time evolution in dark matter, baryon fraction $f_b$, metal mass, and metallicity respectively within the virial radius of the most massive progenitor of these four halos.  Black circles represent instances of new SNe, with sizes scaled based on the number of new SNe.  Magenta points indicate merger events.

All four halos show a similar dark matter mass evolution, originating very early on, within 50 Myr of each other and maintaining masses within 1 order of magnitude.  Despite their similar dark matter evolution, they have widely varying gas evolution, with baryon fractions at times differing by up to 5 orders of magnitude.  Rapid early decreases in ${f_b}$ are associated with the explosion of the first SNe.

Halos 1 and 2, which reach the lowest baryon fractions, have among the lowest dark matter masses at the minimum baryon fraction, hinting that the retention of material is tied to the depth of the potential well of the halo, which is closely tied to $T_{\rm vir}$ and is redshift dependent.  At their $f_b$ minima, Halos 1 and 2 have $T_{\rm vir} \approx 798$ K ($v_c \sim 4.7\, \rm km\,s^{-1}$) and $1056$ K ($v_c \sim 5.4\, \rm km\,s^{-1}$) respectively, indicating shallower potentials than Halos 0 and 3, which have $T_{\rm vir} \approx 3295$ ($v_c \sim 9.5\, \rm km\,s^{-1}$) and $2800$ K ($v_c \sim 8.8\, \rm km\,s^{-1}$) respectively.  Although Halo 0 has a similar dark matter mass to the other halos at the time of its first SN, it is in a stage of rapid dark matter growth, which slows the loss of gas and brings in new gas, keeping the baryon fraction from dropping more than an order of magnitude.

The time evolution of metal mass also demonstrates interestingly similar characteristics between these four halos.  All halos, at the time of their first SNe, show a rapid spike in their metal mass, immediately followed by a metal loss of 2--3 orders of magnitude.  Similar characteristics can be seen in the metallicity evolution.  Subsequent SNe over the next $\sim$ 25 Myr also result in the loss of metal mass, after which halos begin to steadily acquire metals and gas from accretion, SNe, and winds.  Most importantly, SN explosions after this point no longer appear to drive metal mass out of the halo, although they also no longer appear to contribute greatly to the total metal mass of the halo. This may indicate that, while the bulk of the metal mass, which is being accreted, is not being expelled from the halo, SNe yields are still escaping the halo.  This is supported by the flat (or in the case of Halo 2, decreasing) metallicity with time, indicating that the ratios of metals to gas are not increasing as would be expected if yields injected by SNe were fully retained. As a result, metallicities only begin to increase near the end of the simulation.

\subsubsection{Transport of Metals from the Halo}
Figure \ref{fig:metloss_ev} shows the evolution of the metal loss fraction $\ell$ with $M_{\rm dm}$.  All four halos show the same characteristic shape in the evolution of $\ell$, albeit with variations, such as the metal injection event in Halo 0 at a mass of $\sim 3 \times 10^6 \, \rm M_\odot$, that likely depend on the time spacing of the chosen simulation outputs.  Using more tightly spaced simulation outputs would likely reveal more of these variations. Figure \ref{fig:metloss_ev} shows that, following the first SNe, halos lose nearly all of the metals produced by stars within their virial radius.  All four halos remain in this state of metal loss until they attain a mass of $\sim 10^7 \, \rm M_\odot$. These halos then enter a period of rapid accretion whereby nearly all the metal mass produced returns to the halo (although Halo 2 does not explicitly show this, the last time step in our simulation shows a slight downturn that suggests it may be entering this stage).  It is only at these high halo masses that halos no longer experience significant loss of the metals (or gas, see Figure \ref{fig:halo_ev}) because of the injection of energy from new SNe.  
 
We use Figure \ref{fig:met_prof} to demonstrate that the reaccretion of metals is in fact the dominant contributor to the turnover in the metal loss fraction evolution.  Figure \ref{fig:met_prof} shows the evolution of the metal mass profiles for Halos 0 and 2, chosen for the differences in their evolution.  If there were no evolution interior to the virial radius, and metal mass only increased via the smooth accretion of metals from the IGM, we would expect a smooth increase in metal mass from the outside in over time.  However, a number of distinct features are apparent.  In both halos, a sudden increase in metal mass near the center of the halo is indicative of an SN injecting metals into the ISM.  Subsequently, the metal mass profile relaxes, with metals in the center moving toward and beyond the outskirts of the halo. However, as the halo continues to evolve, the metal mass begins to build up again, from the outside in, indicating accretion of metals.  Thus, the reaccretion of metals is a significant contributor to the total amount of metals in each halo.  While Halo 2 shows fairly simple evolution via smooth accretion following the first SNe, Halo 0 has a more complex evolution.  Near the final redshift (yellow lines), we see that SNe once again evacuate metals from the center of the halo.  However, in this case, the metals are not removed beyond the virial radius, an indicator that the potential well has grown deep enough for the halo to retain its metals.
 
While we argue that this is a mass-dependent phenomenon and a result of the depth of the potential well, it is important to note that the point at which halos retain metals may also be correlated with time, as time and halo mass growth are not independent of one another.
 
Notably, our results are similar to those of \citet{Kitayama2005}, who demonstrated that, in low-mass halos ($M_{\rm dm} <10^6 \, \rm M_\odot$), SN explosions, even with low input energy, lead to the complete evacuation of a halo's gas in the presence of radiative feedback that lowered gas densities.  Meanwhile, they found that, in higher-mass halos ($M_{\rm dm} \sim 10^7 \, \rm M_\odot$), gas densities remain high and confine the expansion of the shock.  \citet{Hicks2021} also found that halos $M_{\rm dm} < 10^6 \, \rm M_\odot$ do not retain their metals, and in fact, most of the metals eventually incorporated in star formation originated beyond the virial radius of the halo. \citet{Muratov2013} conducted an analysis similar to our study, and found that 80\% of metals were retained within the virial radius for halos with $M_{\rm dm} >3\times10^6 \, \rm M_\odot$ at the first pair-instability SN event.  We note that their analysis is not exactly comparable as the \citet{Muratov2013} simulations only resolve star formation in halos with $M_{\rm dm} > 10^6 \, \rm M_\odot$, whereas \Aeos is able to resolve star formation in minihalos with $M_{\rm dm} > 10^5 \, \rm M_\odot$, implying that nearly all halos in \Aeos that form stars have their first SNe at a mass of  $M_{\rm dm} < 3\times10^6 \, \rm M_\odot$ (see Figure \ref{fig:first_popiii}).

\begin{figure*}
    \centering
    \includegraphics[width=0.95\linewidth, trim=4.15cm 6.55cm 6cm 8.2cm, clip]{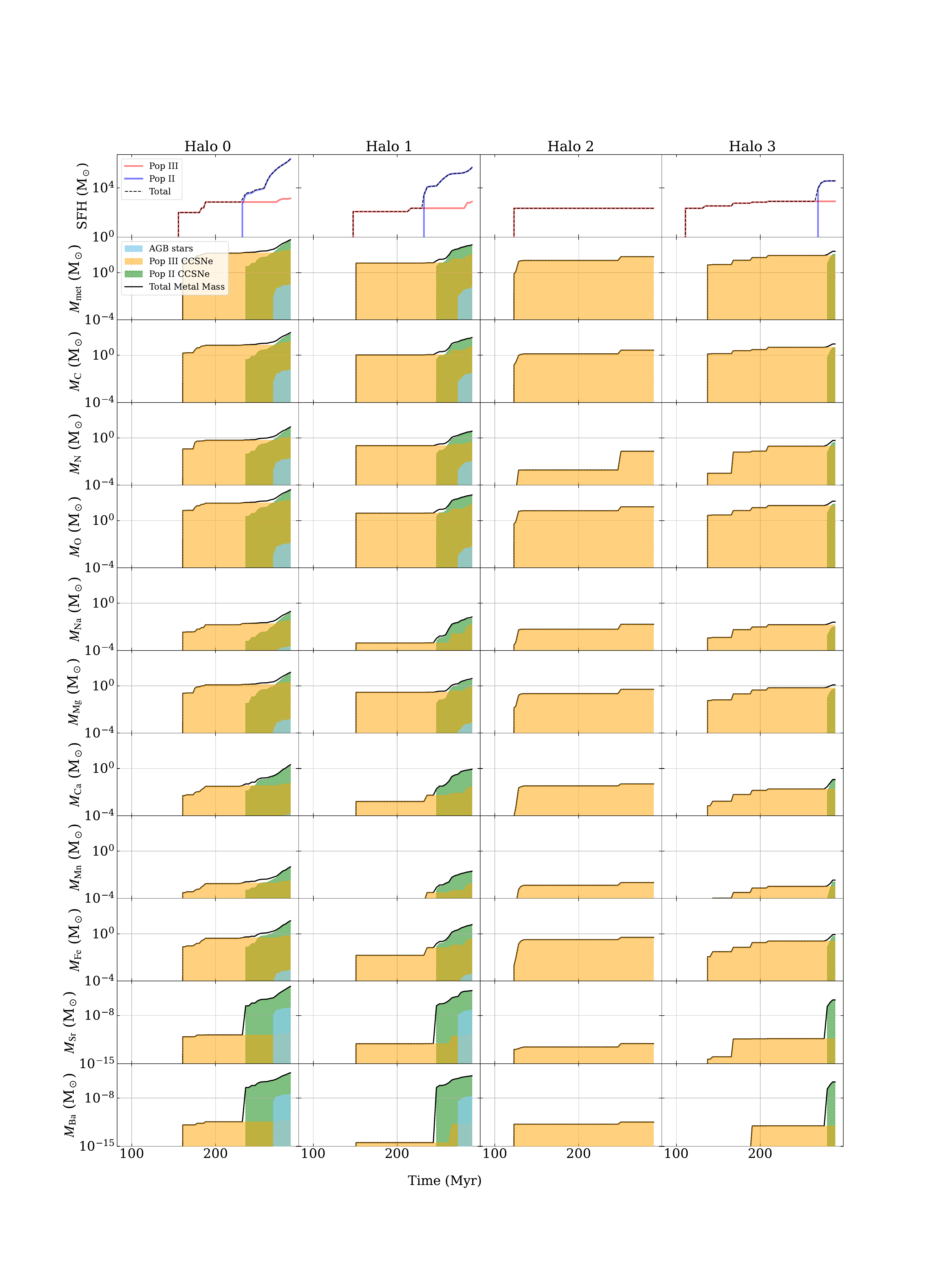}
    \caption{Total injected yield ({\em black line}) for all metals as well as individual elements from AGB winds ({\em cyan}) and Pop III ({\em orange})/Pop II ({\em green}) CCSNe occurring within $r_{\rm vir}$, for each of the four most massive halos over time.  Note the different y-scale for elements Sr and Ba. The top panel shows the star formation history of the main progenitor line for each halo.}
    \label{fig:nuc_comp}
\end{figure*}

\subsubsection{The Extent of Enrichment beyond $r_{\rm vir}$}
Given that halos lose nearly all metals early in their evolution, an interesting question to ask is how far are metals transported, and at what point are they retained in the central, star-forming region of the halo. Figure \ref{fig:frac_rvir} shows the evolution of $\ell$ within various multiples of $r_{\rm vir}$ from 0.5$r_{\rm vir}$ out to 2$r_{\rm vir}$.  We do not explore $\ell$ beyond 2$r_{\rm vir}$ as these halos are in overdense regions, and larger multiples of the virial radius will begin to overlap with the virial radii of other halos.  All halos, with the exception of Halo 0, lose metals beyond at least 2$r_{\rm vir}$.  In fact, Halo 0 and Halo 1 are examples of the importance of considering the halo environment. As can be seen in Figure \ref{fig:box_projection}, Halos 0 and 1 share a dense environment, and can readily exchange material.  The black vertical line across Halos 0 and 1 in Figure \ref{fig:frac_rvir} indicates their evolutionary state at 160 Myr, the point at which the first SNe in Halo 0 expels its metals.  This correlates with the point at which Halo 1 begins to acquire metals within 2$r_{\rm vir}$, as the metals leaving Halo 0 are accreted by Halo 1.  Additionally, for Halo 0, the fraction of (injected) metals lost beyond $2r_{\rm vir}$ only reaches about 30\%, likely due to incoming (accreting) metals that were ejected from Halo 1 about 10 Myr earlier.

Although ejection of metals beyond $r_{\rm vir}$ may not be dependent on the number of SNe, ejection at farther distances may be.  Specifically, Halo 2 experiences several more SN explosions early in its evolution compared to Halos 0, 1, and 3.  Immediately following this series of explosions, Halo 2 reaches baryon fractions and metal masses at least 3 orders of magnitude lower than its counterparts (see Figure \ref{fig:halo_ev}), despite having similar initial values.  Figure \ref{fig:frac_rvir} shows that, while Halos 0, 1, and 3 begin to reaccrete metals within 2$r_{\rm vir}$ at halo masses of $\sim 10^6 \,\rm M_\odot$, Halo 2 loses all metals beyond 2$r_{\rm vir}$ until it reaches a mass of $\sim 10^7 \, \rm M_\odot$. In fact, we find that over 80\% of halos have metal enrichment bubbles (containing 50\% of the metals produced) that extend beyond $2r_{\rm vir}$, which, when converted to physical distances, is consistent with the strong feedback model in Figure 2 of \citet{Griffen2018}.

\subsection{Case Study: Intermediate Mass Halos}
For completeness, we also consider a set of seven intermediate mass halos with dark matter masses between $5 \times 10^6$ and $10^7 \, \rm M_\odot$. Figure \ref{fig:metloss_ev_int} shows that each of these halos has a similar characteristic evolution to their more massive counterparts, particularly with regards to the baryon fraction, which displays a nearly monotonic increase in the retention of gas once a halo reaches a characteristic mass of $\sim 10^6\, \rm M_\odot$.  Notably, no clear monotonic decrease in $\ell$ is found, potentially with the exception of the most massive intermediate mass halo, as none of these halos achieve the characteristic mass of $10^7\, \rm M_\odot$ by the final redshift.

\section{Individual Elements}
\label{sec:indiv_elem}
One of the key features of \Aeos is the ability to resolve metal mixing from individual stars on parsec scales.  By tracing a set of elements that are produced through different nucleosynthetic channels, we can uncover the contributions of each channel to the enrichment of each galaxy and its environment.

\subsection{Nucleosynthetic Contributions to Individual Elements}

Given the yield table for \Aeos and the set of stellar remnants that have produced Pop III or Pop II CCSNe, or have gone through an AGB phase, we can sum up the total injected metal mass as well as the injected metal mass of each individual element that each nucleosynthetic channel has contributed to their host halo.  Figure \ref{fig:nuc_comp} shows the breakdown of metal mass (total and individual elements) contributed by each channel to the four most massive halos in \Aeos.  The top row shows the cumulative star formation history broken down by Pop III and Pop II stars for each halo, enabling a comparison of the total stellar mass to the contribution in each channel.  The remaining rows demonstrate that, for each halo with a Pop II CCSNe contribution, the nucleosynthetic yields from Pop II quickly dominate the Pop III yields once Pop II CCSNe begin.  However, with the exception of the neutron-capture elements (Sr and Ba), Pop II CCSNe dominate the nucleosynthetic production by only 1 or 2 orders of magnitude, despite the nearly 3 orders of magnitude dominance in mass for Pop II stars, due to the mass contribution from low-mass stars.  Figure \ref{fig:nuc_comp} also highlights the wide diversity in star formation and enrichment histories seen across even just the four most massive halos, ranging from little to no impact from AGB stars, or even no component from Pop II stars at all.

We also find that AGB stars do not contribute much to early metal production due to the longer lifetimes of low-mass stars.  Even so, as the primary channel for s-process element production, their contribution to Sr and Ba is significant compared to other lighter elements.  Although AGB stars contribute as much to Sr and Ba as Pop II CCSNe in our yield table (see Appendix of \citetalias{AeosMethods}), we expect that, at later times in the simulation, the AGB contribution to heavy elements will exceed the Pop II CCSN component as the result of a changed initial mass distribution of the available stars.

\subsection{Loss Fraction of Individual Elements}
As in Section \ref{sec:met_loss}, we also examine how individual elements are ejected from their host halos at early times.  This work is motivated by the results from \citet{Emerick2018b}, which demonstrated that elements predominantly ejected by AGB winds were preferentially retained within the disk of an isolated galaxy due to the lower injection energy into the cold neutral ISM compared to the higher energy injection through SNe.

\begin{figure*}
    \centering
    \includegraphics[width=\linewidth, trim=0cm 0cm 6cm 0cm, clip]{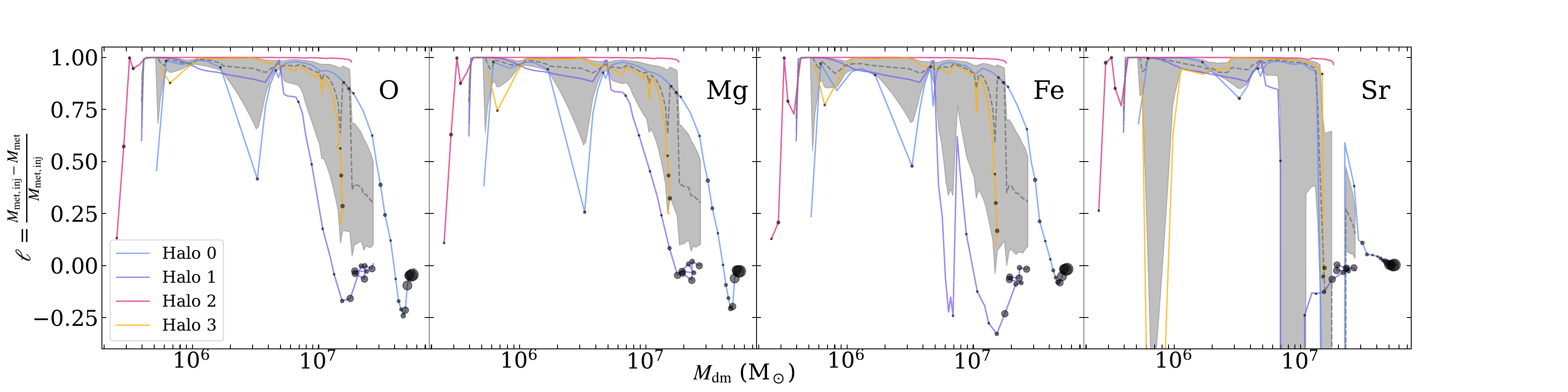}
    \caption{Metal loss fraction $\ell$ within $\rm{r_{vir}}$ of individual elements as labeled, representing light (O), $\alpha$ (Mg), Fe-peak (Fe), and heavy (Sr) elements for the four most massive halos. Lines and symbols are the same as in Figure \ref{fig:metloss_ev}.}
    \label{fig:indiv_elem}
\end{figure*}

To this end, we calculate $\ell$ for all individual elements traced for the four most massive halos in \Aeos.  Figure \ref{fig:indiv_elem} shows the metal loss fraction for one element from each nucleosynthetic group: O (light), Mg ($\alpha$), Fe (Fe-peak), and Sr (s-process).  It is important to note here that there are regions, particularly in Sr and Ba (not shown), that appear to have extreme external enrichment.  We believe this is an analytic issue with our accounting for the injection of AGB winds in our calculations of $\ell$ \textit{after} they are actually injected into the simulation, causing the metal mass contained in the halo to appear to increase before the injected metal mass does.  With the small amounts of Sr and Ba that are produced by each nucleosynthetic event (see Figure \ref{fig:nuc_comp}), it is easy to have an increase in the metal mass of multiple orders of magnitude, causing extreme variation in $\ell$.  Correcting for this would in all likelihood result in an evolution in the metal loss fraction that is similar to that of the other individual elements.

Examining the representative elements shown in Figure \ref{fig:indiv_elem}, they share the same distinct features as the overall $\ell$ for all metals.  This is true for all the elements that we follow, including those not shown. Specifically, we find that $\sim$100\% of individual metals are lost beyond the virial radius of the halo until the halo reaches a mass of $\sim10^7 \, \rm M_\odot$, and that this does not vary between elements from different nucleosynthetic families, including s-process elements produced by AGB stars.  This is unsurprising given Figure \ref{fig:nuc_comp}, which shows that, although AGB contributions to Sr and Ba are significant in comparison to their contributions to other elements, the contribution from CCSNe still dominates the production of Sr and Ba at these times.  Thus, it is expected that we should see the same behavior for all elements at early times.  

If the simulation were run to lower redshifts, we might expect to find that the loss fraction near the center of the halo is significantly lower for elements dominantly produced by AGB stars due to their lower injection energy.  In this case, it could be expected to also find this effect in the average abundances of each element within and outside of the virial radius.  While running to lower redshift will be difficult with a full cosmological simulation, this is an ideal question to study with zoom-in simulations, which we plan to do with \Aeos.

\section{Conclusions}
\label{sec:conclusions}

\subsection{Summary}
In this work, we use our new star-by-star simulation of cosmological galaxy formation, \Aeos, to analyze how individual stellar feedback drives the first steps of heavy element enrichment throughout the study volume as well as in and around individual halos.  We further study the impact from different nucleosynthetic sources on the evolution of individual element enrichment.  Our results show that at early times most of the metals produced are ejected from their parent halos.  Specifically, we find the following:

\begin{enumerate}
    \item Only $\lesssim8$\% of metal mass resides in halos until $z \approx 15.3$, at which point the most massive halos begin to retain metals. Those halos retain $\sim$92\% of all metals contained within halos (31\%--35\% of total metals) at the end of our simulation at $z = 14.3$.
    \item Pop III star formation begins in halos with masses ranging from $10^5$ to $2\times 10^6 \, \rm M_\odot$. At early times, these values are only a factor of 2--3 above the critical mass $M_{\rm crit}$ but within the 0.2 dex scatter derived by \citet{Kulkarni2021} for the streaming velocity $v_{\rm bc} = 0$ with a UV background strength of $J_{\rm LW} \sim 0-1 \, J_{21}$.
    \item Our most massive halos generally have similar evolution of their dark matter mass, baryon fraction, metal mass, and metallicity. In particular, the evolution of metallicity with time does not monotonically increase in minihalos following the injection of metals by the first SNe.  The total metallicity initially decreases following the first SNe as metal mass leaves the halo, and does not increase with further enrichment events until metals begin to be reaccreted to and retained within the halo.
    \item Small numbers of halos can be found in transient, metal-rich states immediately after SNe, before the ejecta have had time to escape the halo.
    \item Halos lose all gas and metals to beyond $r_{\rm{vir}}$ after SN enrichment events, regardless of the number of SNe that occur, until they reach a dark matter mass of about $3 \times 10^6$ and $1 \times 10^7 \, \rm M_\odot$ for gas and metals, respectively. The separate characteristic masses for gas and metals suggest that metals in hot shocked ejecta do not efficiently mix with the cooler local gas into which they are injected.
    \item The metal loss fraction $\ell$ decreases with increasing distance beyond $r_{\rm{vir}}$.  For three of the four most massive halos, all metals are expelled beyond at least $2r_{\rm{vir}}$. This is of particular interest for Halos 0 and 1, which have a common environment and pollute each other.
    \item Yields from Pop III and Pop II CCSNe dominate metal production at early times, with the exception of Sr and Ba, for which AGB winds dominate production over Pop III CCSNe yields.
    \item Each individual element follows the same loss fraction behavior as the total metal mass, only retaining metals within the virial radius for halos with dark matter mass above 10$^7 \, \rm M_\odot$.
\end{enumerate}

\subsection{Robustness to Simulation Choices}
As with all simulations, the \Aeos simulations make several choices that potentially influence the robustness of our results.  We detail the effect that some of these choices may have below.

\begin{enumerate}
    \item Altering the IMF may affect quantities such as the relative contribution of Pop III CCSNe to the enrichment history of minihalos, the metal floor [X/H] following Pop III enrichment, and the number of star forming halos due to varying strength of local ionizing sources.  However, due to the independence of the metal loss fraction from the number of SNe exploding in the virial radius of the halo, we do not expect the characteristic mass of metal retention to change so long as Pop III CCSNe are still prevalent among halos. 
    \item If pair-instability SNe were included in our simulations, we might expect that the characteristic masses for gas and metal retention would increase due to the typical injection energy of $10^{53}$ erg, 2 orders of magnitude stronger than a single CCSN.  An instantaneous energy injection of this magnitude could serve to blow metals and gas out farther than the typical CCSN, thereby increasing the halo mass at which they are reaccreted (as seen by the analysis of the SNe closely clustered in time in Halo 2) or to completely unbind the metals and gas from the original halo.  This could result in a greater number of externally enriched halos and a greater extent of enrichment in the IGM.
    \item The dark matter streaming velocity $v_{\rm bc} = 0$ in \Aeos.  However, several studies have demonstrated that the inclusion of streaming velocities can increase the critical mass of halo collapse \citep{Kulkarni2021,Schauer2021,HegdeFurlanetto2023}, reduce the abundance of small halos at high redshift \citep{Naoz2012}, reduce the baryon overdensity in halos \citep{OLearyMcquinn2012,Naoz2013,Conaboy2023}, suppress star formation \citep{Stacy2011,Greif2011b,OLearyMcquinn2012,Conaboy2023,Williams2024}, and increase gas turbulence and thereby fragmentation, affecting the IMF \citep{Greif2011b}.  While we do not expect an altered IMF to affect our results, an increased $M_{\rm crit}$ and the delayed onset of Pop III star formation may.  Although, for reasonable values of $v_{\rm bc}$ \citep[i.e. less than 2$\sigma$, see][]{Kulkarni2021}, star formation is still expected in minihalos $M_{\rm dm} < 10^7 \, \rm M_\odot$, the onset of Pop III star formation will occur in larger halos, which may suppress the extent of the outflows from SNe feedback.  Specifically, halos first forming stars at $M_{\rm dm} > 3\times10^6 \, \rm M_\odot$, the critical mass for gas retention, may no longer expel gas following SNe.  This could result in the metallicity of the halo monotonically increasing in time.  For reasonable values of $v_{\rm bc}$, we may also still expect that the characteristic mass of metal retention remains $M_{\rm dm} \sim 10^7 \, \rm M_\odot$.  For halos forming stars $M_{\rm dm} > 3\times10^6 \, \rm M_\odot$, this will reduce the time that Pop III star formation can occur following an enrichment event.
\end{enumerate}

\subsection{Implications for Numerical and Semi-Analytic Models}
Our findings challenge the classical view that metal enrichment in halos increases monotonically with time and number of SNe.  This has implications particularly for semi-analytic models of the transition from Pop III to Pop II star formation such as \citet{Visbal2020}, which assume that halos retain the SN ejecta so that metal mass and metallicity increase monotonically with time, or \citet{ASLOTH}, which does not account for differences in metallicity between SN-driven outflows and the average ISM gas, which would underestimate the metallicity of the outflow.  Such assumptions can result in early triggering of enriched star formation, and miss the potential for extended Pop III star formation.  Early triggering of enriched star formation may also result in underestimating the metallicity of the first generations of enriched stars due to factors in the halo's environment such as external enrichment and the mixing of metals in the IGM that then reaccrete to the halo.

Through its star-by-star treatment of star formation, \Aeos is poised to study the chemical abundance patterns that result from the mixing of mass-dependent SN yields in both the ISM and IGM.  The imprints of these patterns on the first enriched Pop II stars are observable, and the implications of these signatures for the conditions of the early Universe will fall to star-by-star cosmological simulations to interpret. Going forward, \Aeos will be able to contribute to answering questions such as what is the metal floor for enriched star formation (\citetalias{AeosMethods}), on what scale do SN yields mix in the early Universe, how many SN abundance patterns are represented in the abundances of the first enriched stars, and how do enriched stellar abundances deviate from the local gas abundances and the predictions of homogeneous mixing models.

\section*{Acknowledgments}
The authors wish to thank the anonymous referee for a thorough review that has improved the clarity and completeness of the manuscript.

J.M. acknowledges support from the NSF Graduate Research Fellowship Program through grant DGE-2036197. K.B. is supported by an NSF Astronomy and Astrophysics Postdoctoral Fellowship under award AST-2303858. G.L.B. acknowledges support from the NSF (AST-2108470 and AST-2307419, ACCESS), a NASA TCAN award, and the Simons Foundation through the Learning the Universe Collaboration. This research was supported in part by grant NSF PHY-2309135 to the Kavli Institute for Theoretical Physics (KITP).  M.-M.M.L. and E.P.A. acknowledge partial support from NSF grant AST-2307950. A.P.J. acknowledges support by NSF grants AST-2206264 and AST-2307599. J.H.W. acknowledges support by NSF grant AST-2108020 and NASA grant 80NSSC21K1053. E.P.A. also acknowledges support from NASA ATP grant 80NSSC24K0935. A.F.\ acknowledges support from NSF grant AST-2307436.

The authors acknowledge the Texas Advanced Computing Center at The University of Texas at Austin for providing HPC and storage resources that have contributed to the research results reported within this paper.

\bibliographystyle{yahapj}
\bibliography{refs}

\end{document}